\documentclass[12pt]{amsart}
\usepackage{amsmath,amssymb}

\usepackage{epic,eepic}

\renewcommand{\phi}{\varphi}

\begin{document}

\title{Separation of scattering and selfaction revisited}
\author{L.~D.~Faddeev}
\address{St.Petersburg Department of Steklov Mathematical Institute}

\begin{abstract}
    The definition of scattering operator in Quantum Field Theory is
    critically reconsidered. The correct treatment of one-particle states
    is connected with separation of selfaction from interaction.
    The formalism of functional integral is used for the description
    of such a separation via introduction of the quantum equation
    of motion.
\end{abstract}

\maketitle

    When I got invitation to contribute to the volume, dedicated to
    anniversary of Lev Lipatov I immediately agreed.
    One reason is evident --- I highly respect Lev for his dedication to
Quantum Field Theory, deep intuition and fantastic technical skill.
    The second is the planned title of this volume --- ``Subtleties of QFT''.
    So I decided to popularize one aspect of QFT which is not
    widely known. It is connected with definition of scattering.

    When I was learning QFT in the end of 50-ties of previous century,
    the scattering operator was defined by famous formula
\begin{equation*}
    S = \lim_{\substack{t''\to\infty \\ t'\to -\infty}}
	e^{iH_{0}t''} e^{-iH(t''-t')} e^{-iH_{0}t'}
\end{equation*}
    with introduction of the interaction representation together with
    adiabatic limit, and use of the
    Wick theorem. Derivation of Feynman diagrams by Dyson
\cite{Dyson}
    via these means was most popular.
    The infinities appearing in calculation were dealt with by renormalization
    and divergence of mass and charge were treated on the same footing.

    With some experience in quantum theory of scattering I was somewhat
    unhappy. Indeed, I already knew, that for the limit such as above
    the continuous spectra of
$ H_{0} $ and 
$ H $
    were to coincide, like it happens in scattering on potential where
\begin{equation*}
    H = -\frac{1}{2m} \Delta + v(x) = H_{0} + V
\end{equation*}
    with 
$ v(x) $
    vanishing at large distances. On the other hand everybody knows,
    that perturbation shifts discrete spectrum.

    In relativistic field theory discrete spectrum appears for the
    one-particle states when we consider subspace of fixed momentum
\begin{equation*}
    P\psi = p\psi .
\end{equation*}
    The eigenvalue of free energy has the form
\begin{equation*}
    H_{0} \psi_{0} = \sqrt{p^{2}+m^{2}} \psi_{0} .
\end{equation*}
    Interaction shifts this eigenvalue: it follows from relativistic
    invariance that corresponding one particle state of full hamiltonian has
    eigenvalue
\begin{equation*}
    H\psi = \sqrt{p^{2}+M^{2}} \psi ,
\end{equation*}
    where
$ M $
    is physical mass of corresponding particle. Thus the mass renormalization
    is not prompted by divergence: it is a necessary step to use the correct
    spectrum.

    In a short note
\cite{LDF}
    I proposed a method to realize the selfaction, leading to construction of
    proper one-particle states before considering scattering.
    Recently I presented this method in my Princeton Lectures
\cite{LDFP},
    directed to mathematicians learning QFT, with rather limited success.

    In
\cite{LDF}
    I considered many body Hamiltonian of general form
\begin{multline*}
    H = \int \omega(k) a^{*}(k) a(k) dk + \\
	+\sum_{m,n} \int v_{mn} (k_{1}\ldots k_{m}, k_{1}'\ldots k_{n}')
    a^{*}(k_{1}) \ldots a^{*}(k_{n}) a(k_{1}')\ldots a(k_{n}') \\
	\delta(k_{1}+\ldots +k_{m}-k_{1}'\ldots -k_{n}')
    dk_{1}\ldots dk_{m} dk_{1}' \ldots dk_{n}' .
\end{multline*}
    Terms of the type
$ v_{m0}, m=1,\ldots $,
$ v_{m1}, m=2,\ldots $
    shift vacuum state
$ \Omega $
    and one particle states
$ a^{*}(k)\Omega $
    of perturbed hamiltonian. My proposal was to consider
    transformed hamiltonian
\begin{equation*}
    H_{\text{scatt}} = R H R^{-1} ,
\end{equation*}
    where
$ R $
    is chosen to cancel the dangerous terms mentioned above.
    In calculation of
$ R $
    one encounters denominators of type
$ \sum_{l}\omega(k_{l}) $ and
$ \sum_{l}\omega(k_{l})-\omega(k) $,
    which do not vanish if the condition of stability
\begin{equation*}
    \omega(k_{1}) + \omega(k_{2}) > \omega(k_{1}+k_{2})
\end{equation*}
    is imposed. Thus no zero denominators, leading to imaginary parts,
    appear.

    The alternative interpretation of this trick is that one uses the same
$ H $,
    but with operators
\begin{equation*}
    b(k) = R a(k) R^{-1} , \quad
    b^{*}(k) = R a^{*}(k) R^{-1}
\end{equation*}
    defining the representation of canonical commutation relations,
    different from original ones. In terms of these operators the
    Hamiltonian acquires the form
\begin{equation*}
    H = \int \hat{\omega}(k)b^{*}(k) b(k) dk +
	\sum_{m,n \geq 2} \hat{V}_{mn} ,
\end{equation*}
    where one-particle energy
$ \hat{\omega}(k) $
    differs from
$ \omega(k) $
    and terms
$ V_{m0}$, $V_{0n}$, $V_{m1}$, $V_{1n} $
    in interaction are absent.
    For such Hamiltonian the scattering operator
$ S $
    from the above definition exists.

    The main defect of my method was its nonrelativistic nature.
    Here I use an opportunity to deliver an alternative approach which is
    manifestly Lorentz invariant. No doubt I shall use the Feynman functional
    integral and treat it in a variant of the background field method
\cite{DeWitt},
\cite{Hooft},
\cite{ArFSl}.

    The 
$ S $-matrix is defined as a limit of the transition operator
\begin{equation*}
    U(\phi''(x'',t''),\phi'(x',t')) = \int \exp \frac{i}{\hbar}
	S_{t'}^{t''}(\phi) \prod_{x,t'<t<t''} d\phi
\end{equation*}
    with prescribed asymptotics of initial and final configurations when
$ t' \to -\infty $,
$ t'' \to \infty $.

    Here
$ S_{t'}^{t''}(\phi) $
    is corresponding action functional. Calculations in the background method
    begin with ansatz
\begin{equation}
\label{phia}
    \phi = \phi^{\text{ph}} + \sqrt{\hbar} \chi ,
\end{equation}
    where 
$ \phi^{\text{ph}} $ and
$ \chi $
    satisfy the asymptotic conditions and appropriate radiation condition,
    correspondingly.

    Usually
$ \phi^{\text{ph}} $
    is taken to satisfy the classical equations of motion
\begin{equation*}
    \frac{\delta S}{\delta \phi} \Bigr|_{\phi=\phi^{\text{ph}}} = 0
\end{equation*}
    with given asymptotic conditions. This would be in spirit of adiabatic
    approach, used in Dyson method.
    My main statement is that this proposal is too naive and should be
    modified to take into account selfaction.

    To make my proposal more explicit consider the case of scalar field
$ \phi(x) $
    with action
\begin{equation*}
    S=\frac{1}{2} \int [(\partial_{\mu} \phi)^{2} - m^{2}\phi^{2} -V(\phi)]dx
\end{equation*}
    (Since I do not believe in the nontriviality of the most popular case
$ V(\phi) = \lambda\phi^{4} $ for
$ \lambda>0 $,
    I take this example only for formal illustration; another possibility is
    to take case of YM, as it is discussed in
\cite{LDC},
    but vector indexes will distract us from the main point).

    Under ansatz 
(\ref{phia})
    we get
\begin{align*}
    \frac{1}{\hbar} S(\phi) = & \frac{1}{\hbar} S(\phi^{\text{ph}}) +
	\frac{1}{\sqrt{\hbar}} \int V_{1} (\phi^{\text{ph}}) \chi(x) dx +\\
    &+ \frac{1}{2} \int (\chi(x)M\chi(x)) dx
	+ \sqrt{\hbar} \int V_{3}(\phi^{\text{ph}}) \chi^{3} dx + ... ,
\end{align*}
    where
\begin{equation*}
    V_{1}(\phi^{\text{ph}}) = (\Box + m^{2})\phi^{\text{ph}} +
	\frac{\delta V}{\delta\phi}\Bigr|_{\phi=\phi^{\text{ph}}}
\end{equation*}
    is LHS of classical equation of motion, linear differential operator
$ M $,
    defining the quadratic form, is given by
\begin{equation*}
    M\chi = \Box\chi +m^{2}\chi 
	+ \frac{\delta^{2}V}{\delta\phi^{2}}
	\Bigr|_{\phi=\phi^{\text{ph}}} \chi
\end{equation*}
    and
$ V_{3} $ etc.
    are given by higher derivatives of
$ V $ at
$ \phi=\phi^{\text{ph}} $.

    The Gaussian formal calculation of the functional integral gives
\begin{equation*}
    U = \exp\bigl\{ \frac{i}{\hbar}S(\phi^{\text{ph}})
	- \frac{1}{2} \ln\det M + \sum \text{closed diagrams} \bigr\} ,
\end{equation*}
    where diagrams are constructed via vertices
\begin{equation*}
    V_{1} =
\begin{picture}(40,30)
    \drawline(5,2)(35,2)
    \drawline(1,4)(5,0)
    \drawline(1,0)(5,4)
\end{picture} , \quad
    V_{3} = 
\begin{picture}(40,30)
    \drawline(20,4)(20,19)
    \drawline(20,4)(9,-9)
    \drawline(20,4)(31,-9)
\end{picture} , \quad
    V_{4} =
\begin{picture}(40,30)
    \drawline(6,-9)(34,19)
    \drawline(6,19)(34,-9)
\end{picture}
\end{equation*}
    connected by line
$ G(x,y) = 
\begin{picture}(22,15)
    \drawline(0,3)(18,3)
\end{picture} $,
    which is a Green function of operator
$ M $, uniquely defined due to the appropriate radiation conditions,
    imposed on
$ \chi $.
    The first examples are
\begin{equation*}
\begin{array}{ccccc}
\begin{picture}(30,15)
    \drawline(0,5)(25,5)
    \drawline(-4,7)(0,3)
    \drawline(-4,3)(0,7)
    \drawline(25,7)(29,3)
    \drawline(25,3)(29,7)
\end{picture} &
\begin{picture}(50,20)(-5,10)
    \drawline(5,15)(29,15)
    \drawline(1,17)(5,13)
    \drawline(1,13)(5,17)
     \put(40,15){\ellipse{20}{20}}
\end{picture} & 
\begin{picture}(50,20)(-15,10)
    \drawline(1,15)(19,15)
    \put(10,15){\ellipse{20}{20}}
\end{picture} &
\begin{picture}(40,20)(0,5)
    \qbezier(15,10)(29,0)(29,10)
    \qbezier(15,10)(29,20)(29,10)
    \qbezier(15,10)(1,0)(1,10)
    \qbezier(15,10)(1,20)(1,10)
\end{picture} &
\begin{picture}(60,20)(-10,10)
    \drawline(7,15)(29,15)
     \put(-4,15){\ellipse{20}{20}}
     \put(40,15){\ellipse{20}{20}}
\end{picture}  
 \\
    a & b & c & d & e 
\end{array}
\end{equation*}   

    We shall distinguish weakly connected and strongly connected diagrams.
    The diagrams of the first type can be separated in two by cutting one
    line. The diagrams
$ a $, $ b $, $ e $
    in the fig.1 are weakly connected. It is clear that contribution of
    such diagram is given by
\begin{equation*}
    \int \Gamma_{1}(x) G(x,y) \Gamma_{2}(y) dx \, dy ,
\end{equation*}
    where each factor can be depicted as a diagram with one external line.
    The examples, corresponding to diagrams 
$ a $, $ b $, $ e $
    look as follows
\begin{equation*}
    \begin{picture}(50,20)(0,10)
    \drawline(5,15)(35,15)
    \drawline(1,17)(5,13)
    \drawline(1,13)(5,17)
\end{picture} , \quad
\begin{picture}(60,20)(-10,10)
    \drawline(21,15)(39,15)
     \put(10,15){\ellipse{20}{20}}
\end{picture} .
\end{equation*}
    The first, as was already stated, is the LHS of the classical equation
    of motion. Now we formulate the main proposal: The background field
$ \phi^{\text{ph}} $
    is to be taken as solution of equation
\begin{equation*}
    \begin{picture}(50,20)(0,10)
    \drawline(5,15)(35,15)
    \drawline(1,17)(5,13)
    \drawline(1,13)(5,17)
\end{picture} +
\begin{picture}(60,20)(-10,10)
    \drawline(21,15)(39,15)
\texture{55888888 88555555 5522a222 a2555555 55888888 88555555 552a2a2a
2a555555
    55888888 88555555 55a222a2 22555555 55888888 88555555 552a2a2a 2a555555
    55888888 88555555 5522a222 a2555555 55888888 88555555 552a2a2a 2a555555
    55888888 88555555 55a222a2 22555555 55888888 88555555 552a2a2a 2a555555
}
    \put(10,15){\shade\ellipse{20}{20}}
\end{picture} = 0 ,
\end{equation*}
    where the bubble is sum of strongly connected diagrams.
    I propose to call this condition quantum equation of motion.
    We can presume, that the solution is uniquely defined by asymptotic
    conditions as it was supposed in the case of classical equation of
    motion.
    This must be argued for as now this equation is nonlocal.

    As soon as the
$ \phi^{\text{ph}} $
    is chosen via quantum equation of motion the functional transition
    amplitude acquires the form
\begin{align*}
    \frac{1}{i}\ln U =& W(\phi^{\text{ph}}) 
	= \frac{1}{\hbar} S(\phi^{\text{ph}})
	- \frac{1}{2i} \ln \det M \\
    & + \sum \text{closed strongly connected diagrams}
\end{align*}
    and the last series contains terms of order
$ \sqrt{\hbar} $
    and higher.

    Our receipt explicitly takes into account the selfaction effects.
    In particular quantum equation of motion produce shift of mass for
    free particle, parameterizing the asymptotic behaviour of solutions.

    I finish this methodological exposition by two comments

    1. The role of classical solutions, in particular of instantons,
    should be carefully reconsidered.

    2. Quantum equations could have soliton solutions, which are absent
    in the classical limit. In particular, it is not completely crazy
    idea that quantum Yang-Mills equations have soliton-like solutions
    due to the dimensional transmutation. The concrete proposal for
    that are discussed in
\cite{LDN}--\cite{LD}.

\end{document}